\documentclass[10pt]{IEEEtran}

\newcommand{\camerareadyonly}[1] {#1}

\usepackage[T1]{fontenc}

\usepackage{fancyhdr}
\usepackage{booktabs} 
\usepackage{listings}
\usepackage[]{algorithm2e}
\usepackage[]{graphicx}
\usepackage[]{caption}
\usepackage{subcaption}
\usepackage{textcomp}
\usepackage[cmex10]{amsmath}
\usepackage{amssymb}
\camerareadyonly{
\usepackage[symbol]{footmisc}
}
\usepackage{mathtools}
\usepackage{fixltx2e}
\usepackage{url}
\usepackage{color}

\newcommand{\blockcomment}[1] {}
\newcommand{\shell}[1]{\texttt{#1}}
\newcommand{\bfit}[1]{\textit{#1}}

\definecolor{codebgcolor}{rgb}{0.95,0.95,0.95}
\newcommand{\ourapproach}{\textsc{TensorTuner}}
\newcommand{\intel}{{Intel\!}~\textsuperscript{\textregistered}}
\newcommand{\xeon}{{Xeon\!}~\textsuperscript{\textregistered}}
\newcommand{\xeonphi}{Xeon {Phi\!}~\textsuperscript{TM}}

\begin{document}

\title{Auto-tuning TensorFlow Threading Model for\\ CPU
Backend\thanks{Presented at ML-HPC workshop, held along with
SuperComputing'18, Dallas, TX. DOI 10.1109/MLHPC.2018.000-7}}

\author{
\IEEEauthorblockN{Niranjan Hasabnis\\}
\IEEEauthorblockA{Intel Corporation\\
Santa Clara, CA.\\
niranjan.hasabnis@intel.com}
}

\maketitle

\begin{abstract}

TensorFlow is a popular deep learning framework used by data scientists to solve
a wide-range of machine learning and deep learning problems such as image
classification and speech recognition. It also operates at a large scale and in
heterogeneous environments --- it allows users to train neural network models or
deploy them for inference using GPUs, CPUs and deep learning specific
custom-designed hardware such as TPUs. Even though TensorFlow supports a variety
of optimized backends, realizing the best performance using a backend may
require additional efforts. For instance, getting the best performance from a
CPU backend requires careful tuning of its threading model.  Unfortunately, the
best tuning approach used today is manual, tedious, time-consuming, and, more
importantly, may not guarantee the best performance.

In this paper, we develop an automatic approach, called {\ourapproach}, to
search for optimal parameter settings of TensorFlow's threading model for CPU
backends.  We evaluate {\ourapproach} on both Eigen and Intel's MKL CPU backends
using a set of neural networks from TensorFlow's benchmarking suite.  Our
evaluation results demonstrate that the parameter settings found by
{\ourapproach} produce 2\% to 123\% performance improvement for the Eigen CPU
backend and 1.5\% to 28\% performance improvement for the MKL CPU backend over
the performance obtained using their best-known parameter settings. This
highlights the fact that the default parameter settings in Eigen CPU backend are
not the ideal settings; and even for a carefully hand-tuned MKL backend, the
settings are sub-optimal.  Our evaluations also revealed that {\ourapproach} is
efficient at finding the optimal settings --- it is able to converge to the
optimal settings quickly by pruning more than 90\% of the parameter search
space.

\end{abstract}

\begin{IEEEkeywords}
Auto-tuning; HPC; Deep Learning; Black-box Optimization; Software Optimization;
\end{IEEEkeywords}

\section{Introduction}

Machine learning has seen phenomenal growth in the last few years, and is being
applied to a variety of problems in different fields. This success can be
attributed to availability of large datasets and easy access to
increasingly-powerful computational resources (thanks to Moore's Law) for
processing these datasets.

Realizing the growth in machine learning and deep learning areas, Google
developed and open-sourced TensorFlow \footnote{Other names and brands may be
claimed as the property of others.} \cite{tf-whitepaper, tfgithub} framework
in 2011. TensorFlow has since then become a very popular framework for
developing deep learning models with applications to image recognition, speech
recognition, language translation, NLP, etc. In addition to supporting various
types of models, TensorFlow also supports both large-scale training and
inference: models can be trained on a large distributed cluster of different
devices such as CPUs, GPUs, and TPUs, and inference can be done on a device as
small as a mobile phone \cite{tfmobile}.

TensorFlow uses Eigen \cite{eigen} template library as a default implementation
for its CPU backend. Realizing that the default implementation does not deliver
best training and inference performance on \camerareadyonly{{\intel}
{\xeon}\footnote{Intel, the Intel logo and {\xeon} are trademarks of Intel
Corporation or its subsidiaries in the U.S. and/or other countries.
{\textcopyright} Intel Corporation.}} and {\xeonphi} platforms, Intel
open-sourced an alternative implementation \cite{tfgithub} using its Math Kernel
Library for Deep Neural Networks (MKL-DNN \cite{mkldnn}).  Intel's
implementation delivers up to 70x gains over Eigen CPU backend \cite{inteltf,
inteltfpaper}.

Unfortunately, realizing the best performance even from a highly-optimized
backend like Intel's MKL requires additional efforts. To be precise, TensorFlow
represents a neural network as a data-flow graph and allows users to exploit the
graph-level parallelism by offering a configurable threading model to express
the parallelism. Concretely, TensorFlow's threading model contains following
three parameters: (1) \texttt{inter\_op\_parallelism\_threads}: This parameter
dictates the maximum number of graph nodes that can be executed in parallel, (2)
\texttt{intra\_op\_parallelism\_threads}: This parameter dictates the maximum
number of threads that can be used to execute a graph node, and (3)
\texttt{OMP\_NUM\_THREADS}: This parameter applies to MKL backend only, and it
dictates the maximum number of threads to be used to execute a graph node that
is of type MKL. As one can now realize, the training or inference performance of
a neural network depends on setting the parameters of the threading model
correctly; incorrect settings will not deliver the best possible performance
from a CPU backend.

The threading model parameters can be treated as the hyper-parameters of a
neural network, and one can apply well-researched hyper-parameter tuning
techniques \cite{hyperopt,spearmint,autoweka} to get the best performance.
Unfortunately, tuning techniques used in TensorFlow's CPU backend are not that
advanced --- the default approach used by both Eigen and MKL CPU backends is to
set \texttt{inter\_op\_parallelism\_threads} to a fix number such as 2 or 4 and
to set the remaining two parameters to the available number of CPU cores (by
querying at runtime). It is easy to realize that such an approach does not
deliver best performance as it is neural-network agnostic. It is possible,
nevertheless, to override these defaults by explicitly setting the parameters.
Both TensorFlow's publicly-available performance guide
\cite{tfbenchmarkingguide} as well as Intel's blogs on TensorFlow optimizations
\cite{inteltfv2} offer general advice on selecting parameter values (Intel's
blogs also offer specific parameter values for a set of popular neural networks
such as ResNet50.) Unfortunately, as we will see in the Evaluation section, such
a general advice may not deliver best performance. And the last resort of
exhaustively exploring the parameter search spaces does not generalize as it is
an NP-complete problem (it may work for small search spaces though.) An
automated approach that efficiently finds the optimal parameter settings that
deliver the best performance is highly desirable.

In this paper, we propose an automatic approach, called {\ourapproach}, to
efficiently search for optimal parameter settings of TensorFlow's threading
model for CPU backends. We formulate the problem as a \emph{black-box function
optimization problem} and solve the problem using \emph{Nelder-Mead Simplex}
\cite{NM} gradient-free optimization algorithm. We use a set of neural networks
from TensorFlow's benchmarking suite and evaluate our approach using two
criteria: (1) \emph{Tuning quality} measures the performance of the neural
network with the parameter setting suggested by {\ourapproach}, and (2)
\emph{Tuning efficiency} measures the ability of {\ourapproach} to find the
optimal settings quickly.  Our results demonstrate that the optimal settings
found by {\ourapproach} deliver between 1.5\% to 123\% better performance with
Eigen and MKL CPU backends. Moreover, they also demonstrate that {\ourapproach}
is efficient in finding the optimal settings and can prune the parameter search
spaces by as much as 90\%.

\bfit{Contributions.} This paper makes following contributions:
\begin{itemize}

\item To the best of our knowledge, ours is the first effort to solve the
problem of tuning TensorFlow's threading model for CPU backends.

\item Our evaluations demonstrate that {\ourapproach} can find optimal parameter
settings for MKL and Eigen CPU efficiently (10X more efficient than an exhaustive
evaluation). Moreover, the optimal parameter settings found by {\ourapproach}
deliver 1.5\% to 123\% improvement in performance over the best-known parameter
settings for a set of neural network models from TensorFlow's benchmark suite.

\item More importantly, our evaluations demonstrate the approach proposed by
{\ourapproach} is considerably more efficient and qualitatively better than the
current approach of manually tuning TensorFlow's CPU backend.

\end{itemize}

\bfit{Paper organization.} This paper is organized as follows.
Section~\ref{sec:relwork} covers related work, where as
Section~\ref{sec:background} covers the necessary background to understand the
technique. Section~\ref{sec:design} formulates the problem of tuning
TensorFlow's CPU backend, describes the design and implementation of
{\ourapproach}. Evaluation results for both MKL and Eigen CPU backends are then
presented in Section~\ref{sec:eval}.  Lastly, Section~\ref{sec:future} briefly
mentions future work, and Section~\ref{sec:conclusion} concludes the paper.

\blockcomment{
Exponentially-growing parameter search space simply makes exhaustive exploration
time-consuming. Such an approach may work for \emph{offline tuning}, where a
tuner provides optimal parameter settings to known neural network models (such
as the ones provided by Intel's blog). But the approach will not work for
\emph{online tuning}, where a TensorFlow user is developing a new neural network
model and wants to realize maximum performance from TensorFlow's default CPU
backend option (also called as out-of-box performance). Even in case of offline
tuning, an efficient alternative to exhaustive exploration that quickly finds
optimal parameter settings is highly desirable.
}

\section{Background}
\label{sec:background}

The background provided here is not meant to be a
comprehensive description of TensorFlow --- one may refer to references such as
\cite{tf-whitepaper, tfgithub} for a detailed description of TensorFlow. Rather,
this description is supposed to be a short summary of TensorFlow concepts that
are necessary to understand the optimizations discussed in the next section.

\begin{figure}
\centering
\begin{lstlisting}[language=Python,columns=fullflexible,backgroundcolor=\color{codebgcolor}]
 import tensorflow as tf

 a = tf.Variable(tf.zeros([100]))
 b = tf.Variable(tf.ones([100]))
 c = a + b
 d = a - b
 e = c * d

 s = tf.Session()
 result = s.run([e])
\end{lstlisting}
\caption{TensorFlow Python code for \texttt{e = (a + b) * (a - b)}}
\label{fig:example}
\end{figure}

\begin{figure}
\centering
\includegraphics[scale=0.7]{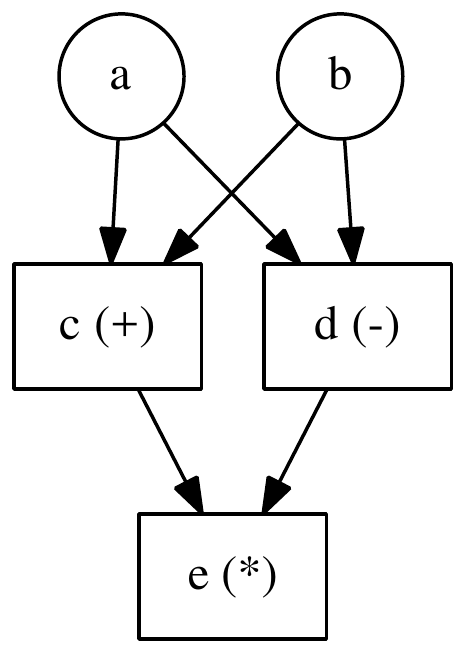}
\caption{Dataflow graph for \texttt{e = (a + b) * (a - b)}}
\label{fig:example_graph}
\vspace{-0.15in}
\end{figure}

\paragraph{\bfit{TensorFlow execution model}} TensorFlow provides a simple
dataflow-based programming abstraction by offering a high-level scripting
interface (typically Python) that allows users (typically data scientists) to
construct dataflow graphs and experiment with them easily.
Figure~\ref{fig:example} shows an example code written in Python to
construct and execute a TensorFlow graph (shown in Figure~\ref{fig:example_graph})
for expression \texttt{e = (a + b) * (a
- b)}. \texttt{a} and \texttt{b} are TensorFlow \texttt{variables}, initialized
  to vector containing one hundred zeros and one hundred ones resp.  Rectangular
graph nodes represent individual mathematical operators (called
\textit{operations}) such as matrix multiplication, convolution, etc, and the
edges between the nodes represent dataflow between the nodes. To be precise,
edges carry multi-dimensional arrays called \textit{tensors}.  After the graph
is constructed, it is executed using TensorFlow's \texttt{Session} APIs.

Execution of a neural network model is mapped to executing operators in the
dataflow graph by considering their constraints. Constraints consist of simple
dataflow constraints (inputs of an operation need to be available to execute it)
and control-flow constraints (a user may explicitly add control-flow constraint
between operations to enforce sequential execution.) When an operator is ready
for execution, TensorFlow runtime calls the \textit{kernel} for that operator on
the device that it has assigned to the operator. Kernel is a function
implementing the operator. A single operation may have multiple registered
kernels with specialized implementations for a particular device or data type.
TensorFlow's CPU backend uses Eigen \cite{eigen} open-source library to
implement CPU kernels for almost all of the TensorFlow operators.

\paragraph{\bfit{Parallel execution}} Given how TensorFlow's dataflow graphs are
executed, it is easy to realize that the dataflow graph shown in
Figure~\ref{fig:example_graph} can execute operators \texttt{+} and \texttt{-}
in parallel. Such a parallel execution enables users to exploit their
hardware-level parallelism to its full potential. Serial execution of \texttt{+}
and \texttt{-}, on the other hand, may waste CPU's computation power by keeping
percentage of CPU idle. TensorFlow offers a threading model with two parameters
to control the parallel execution of its dataflow graph: (1)
\texttt{inter\_op\_parallelism\_threads} specifies the maximum number of
operators that users want to execute in parallel, and (2)
\texttt{intra\_op\_parallelism\_threads} specifies the maximum number of threads
users want to use to execute individual operators. For the example graph shown
above, setting \texttt{inter\_op\_parallelism\_threads} to 2 allows both
\texttt{+} and \texttt{-} to execute in parallel; while setting it to 1
essentially enables serial execution.

\begin{figure}
\centering
\begin{lstlisting}[language=Python,columns=fullflexible,backgroundcolor=\color{codebgcolor},texcl=false]
 session_config = tf.ConfigProto(
   inter_op_parallelism_threads=2,
   intra_op_parallelism_threads=8)
 s = tf.Session(config=session_config)
 result = s.run([e])
\end{lstlisting}
\caption{TensorFlow Python code for setting threading model}
\label{fig:threadingapi}
\vspace{-0.15in}
\end{figure}

TensorFlow offers high-level configuration APIs to specify the values of the
threading model parameters. Figure~\ref{fig:threadingapi} shows the Python code
to set these parameters for the graph shown in
Figure~\ref{fig:example_graph}.

\section{Design and Implementation}
\label{sec:design}

The problem of tuning TensorFlow's CPU backend to deliver maximum performance
for a given neural network is an optimization problem that can be expressed as a
function maximization problem.

\subsection{Problem formulation}

Broadly speaking, performance of TensorFlow's CPU backend is a function of
various types of input parameters such as
\begin{itemize}
\item The neural network along with its hyper-parameters and the input dataset
used for execution,
\item The configuration of the machine used for execution in terms of the
hardware devices and their configurations (micro-architecture of the CPU, cache
sizes, etc), operating system version installed on the machine, the
configuration of the software environment (such as TensorFlow version, Eigen
version, MKL version, compiler version as well as compiler options used to build
CPU backend, Python version, etc).
\end{itemize}

Given such a diverse set of parameters that define the performance of TensorFlow's
CPU backend, precise formulation of the problem requires assuming that a
number of these input parameters are known. To be precise, we concretize the
problem by assuming that:
\begin{itemize}
\item The neural network along with its hyper-parameters used for
execution and the input dataset to the model are known,
\item The configurations of the underlying hardware CPU devices are known,
\item The configurations of the software environment used for running the model
are precisely defined.
\end{itemize}
In other words, assuming that all of these parameters are constant, the performance
$f$ of TensorFlow's CPU backend can be defined as
a function: \[s = f_{C}({\Sigma})\]

Where,
\begin{itemize}
\item $s$ represents performance ``score'' (e.g., images per second is a typical
metric to measure training performance (throughput) of convolutional neural
networks),
\item $C$ represents the set of all the constant parameters discussed above,
\item ${\Sigma}$ represents a set of various parameters of the threading model
for which we are looking for optimal settings. Concretely,
${\Sigma}$ can be defined as: \[\Sigma = \{p_{1}, p_{2}, .., p_{n}\}\] where
$p$ is a parameter, and $n$ is the number of parameters that we are looking
for optimal settings. To be precise, in this paper, we assume that
  \begin{itemize}
  \item ${\Sigma}$ for MKL backend contains the parameters specified in Intel's
blog (i.e., ${\Sigma}$ = \{\texttt{inter\_op}, \texttt{intra\_op},
    \texttt{OMP\_NUM\_THREADS}\})), and
  \item ${\Sigma}$ for Eigen backend contains the parameters from
TensorFlow's threading model (i.e., ${\Sigma}$ =
\{\texttt{inter\_op}, \texttt{intra\_op}\}).
  \end{itemize}
\end{itemize}

Since ${\Sigma}$ is the only input parameter to our performance function, the set of
all possible instantiations of ${\Sigma}$, represented using $\tau$,
represents the parameter search space. Concretely, if $v$ represents the setting
for a parameter $p$, then a single instance of $\tau$ is essentially a map
containing the setting for every parameter of ${\Sigma}$ (It can be represented as
\{$(p_{1}, v_{1}), (p_{2}, v_{2}), .., (p_{n}, v_{n})$\}.) To keep the search
space bounded, we impose strict upper and lower bounds on all the elements of
${\Sigma}$ (i.e., $\forall p \in {\Sigma}, v_{p} \in \{l_{v}, ..., h_{v}\}$,
where $l_{v}$ and $h_{v}$ are the lower and upper bound for $p$.)

Given this formulation, the problem of tuning TensorFlow's CPU backend
for maximum performance can be defined as:
 \[find \quad t \in \tau \mid f(t) > f(t') \, \forall \, t' \in \tau \wedge t \neq t'\]

\subsection{Design}

Since we have formulated our tuning problem as a function maximization problem,
we can use the typical approach of using gradient-based optimizers \cite{NCG,
optimization_book} to solve the problem. Gradient-based optimizers use the
gradient of the objective function to determine the most promising directions
along which one should search. For a large numbers of variables, gradient-based
optimizers are usually the most efficient algorithms.

Unfortunately, gradient-based optimizers have been known to be inefficient at
handling optimization problems with one or more of the following challenges:
non-differentiable functions, mixed variables, multiple local minima, and large
dimensionality. Although the objective function for our tuning problem may not
have all of the above issues (for example, it does not have large number of
variables), it may be non-differentiable, and it definitely has mixed variables.
The key strength of gradient-free methods is their ability to solve problems
that are difficult to solve using gradient-based methods. Furthermore, many of
them are designed as global optimizers and thus are able to find multiple local
optima while searching for the global optimum. As a result, we decided to use
gradient-free optimization method called \emph{Nelder-Mead Simplex} \cite{NM}. A
number of gradient-free methods such as Simulated Annealing, Divided Rectangles
method, Genetic Algorithms, and Particle Swarm optimization have been developed.
But of all of them, Nelder-Mead Simplex is a very simple yet efficient method.
That is why we decided to use it for our approach. Nonetheless, a comparative
analysis and further improvements to {\ourapproach} may be possible.

\paragraph{Nelder-Mead simplex algorithm} Nelder-Mead simplex is a well-known
and popular optimization algorithm for multi-dimensional unconstrained
optimization problems without gradients. Since this algorithm is not the
contribution of this paper, we advise readers to refer to the original paper
\cite{NM} for more details. Since Nelder-Mead simplex is a solution to a
function minimization problem, we convert our function maximization problem into
a minimization problem by using a division function. Specifically, our new
objective function is: \[f'_{C}({\Sigma}) = \dfrac{1}{f_{C}({\Sigma})}\]

With the new objective function, the problem of tuning for maximum performance
can be defined as: \[find \quad
t \in \tau \mid f'(t) < f'(t') \, \forall \, t' \in \tau \wedge t \neq t'\]

\begin{figure}[]
\centering
  \includegraphics[width=0.9\linewidth]{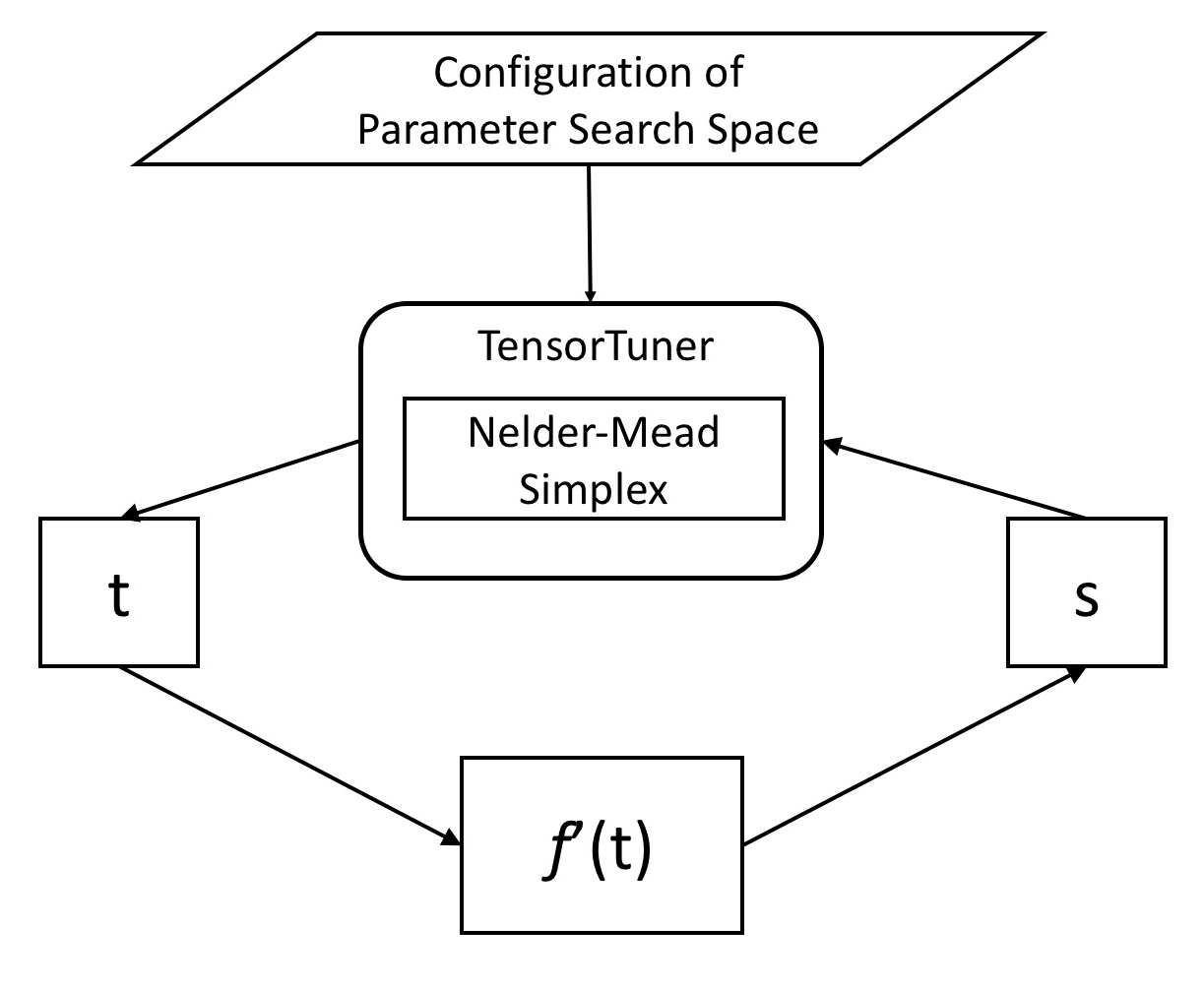}
\caption{{\ourapproach} in operation}
\label{fig:design}
\vspace{-0.15in}
\end{figure}

Figure~\ref{fig:design} shows the design of {\ourapproach}. Given the problem
formulation, {\ourapproach} just needs to produce an assignment for all the
variables of ${\Sigma}$ in every iteration. This assignment is decided by the
Nelder-Mead simplex search strategy by accessing the output of the objective
function for the previous assignment to the variables.  The design of our system
is such that it is very easy to plug-in new search strategies. In addition to
the search strategy, {\ourapproach} also accepts configurations (upper bound,
lower bound and step size) for the variables in ${\Sigma}$. These constraints on
the search space are not necessary --- Nelder-Mead simplex algorithm supports
unconstrained search also. But given that the number of CPU cores are bounded, a
constrained search is likely to converge more quickly.

\subsection{Implementation}

Parameter tuning is a well-known problem in HPC domain. Given that, we utilize
open-source implementation of Active Harmony \cite{AH} tool to implement
{\ourapproach}. Active Harmony project supports infrastructure and algorithms
(such as Nelder-Mead Simplex) to achieve high performance in a distributed,
heterogeneous software systems with changing resource requirements and
capacities. For our implementation, we utilized Active Harmony's
publicly-available source code and wrote a shell script on top
of its \texttt{tuna} tool to accept the configurations for variables and pass
them appropriately to \texttt{tuna}. Specific command lines used to obtain our
evaluation results are mentioned in the Appendix section
(Section~\ref{sec:appendix}).

\section{Evaluation}
\label{sec:eval}

We evaluated effectiveness of {\ourapproach} in terms of tuning TensorFlow's
default Eigen CPU backend and MKL CPU backend. We describe our evaluation
results in terms of: 1) tuning quality, and 2) tuning efficiency.  Tuning
quality is a criterion for comparing search algorithms in terms of their ability
to find the global optimum (and for their ability to avoid local optimums).
Tuning efficiency, on the other hand, evaluates an ability of a search algorithm
to converge to global optimum quickly.

\subsection{Experimental setup}
Before we discuss the results, we discuss the experimental setup used for
the evaluation.

\paragraph{\bfit{Hardware description}} We conducted all of our experiments on
{\intel} {\xeon} Platinum 8180 processor \cite{skl}, running CentOS Linux
version 7.3.1611. We used GCC-6.3.0 for building TensorFlow, and
Python-2.7.5 for running the TensorFlow benchmarks.

\paragraph{\bfit{Eigen backend}} Since Eigen is the default CPU backend for
TensorFlow, we used pre-built TensorFlow-1.7 wheel (the latest version as of
March 2018) for CPU\footnote{\shell{pip install tensorflow==1.7}} for our
experiments.

\paragraph{\bfit{MKL backend}} The source code of Intel-optimized MKL CPU
backend is now present in TensorFlow's github repository \cite{tfgithub}. We
cloned TensorFlow's master branch for commit id
\texttt{7d8ad3da99aba29319c7c9f0e62d567aa2071c21}\footnote{The latest commit
when we started experiments} and built it using the steps listed on the
TensorFlow website \cite{tfbuildwithmkldnn}. We built the wheel for MKL backend
ourselves since the pre-built wheel for version 1.7 was not available from
Intel's webpage \cite{inteltfwheel}.

\paragraph{\bfit{TensorFlow benchmarks}} TensorFlow authors have
open-sourced a suite of popular convolutional neural networks for benchmarking
purpose \cite{tfbenchmarksgithub}. The suite includes popular networks
such as ResNet50, VGG11, InceptionV3 etc.

Table~\ref{table:tfcnn_config} shows the neural network models from TensorFlow's
benchmarking suite that we use for evaluation along with their hyper-parameters
such as batch size and data format. Configurations of batch size and data format
that deliver best performance using MKL CPU backend are publicly-available from
Intel's blog \cite{inteltfv2}. Batch sizes used for Eigen CPU backend evaluation
are same as that of MKL backend evaluation, while the data format is
\texttt{NHWC} (since Eigen CPU backend does not support \texttt{NCHW} data
format.) Rest of the hyper-parameters for both the backends are default
parameters set by TensorFlow's benchmarking scripts.

Intel's blog \cite{inteltfv2} also specifies values for different environment
variables that deliver the best performance on MKL CPU backend. We used these
values for our experimentation and have also documented them in
Table~\ref{table:mkl_tfcnn_config}. Note that Intel's blog does not specify
values for VGG11 and GoogLeNet models. We obtained values of environment
variables for these models from the observation that the values of these
variables are fairly uniform for other models (e.g., \texttt{OMP\_NUM\_THREADS}
is 56 for all other models.) We also confirmed with Intel that the values
that we used were indeed the best-known settings for VGG11 and GoogLeNet.

Our evaluation methodology obtains the performance of TensorFlow's Eigen and MKL
CPU backends with the default settings and compares it with the performance
obtained using optimum settings suggested by {\ourapproach}. We used number of
images processed per second (throughput) as the performance metric for training
and inference scenarios.

\begin{figure}
\centering
\begin{tabular}{||l|l|r|l||}
\hline
\textbf{Backend} & \textbf{Model} & \textbf{Batch} & \textbf{Data} \\
& & \textbf{Size} & \textbf{Format} \\ \hline \hline
MKL CPU & ResNet-50 & 128 & NCHW \\ \hline
MKL CPU & Inception3 & 64 & NCHW \\ \hline
MKL CPU & VGG16 & 128 & NCHW \\ \hline
MKL CPU & VGG11 & 128 & NCHW \\ \hline
MKL CPU & GoogLeNet & 96 & NCHW \\ \hline
Eigen CPU & ResNet-50 & 128 & NHWC \\ \hline
Eigen CPU & Inception3 & 64 & NHWC \\ \hline
Eigen CPU & VGG16 & 128 & NHWC \\ \hline
Eigen CPU & VGG11 & 128 & NHWC \\ \hline
Eigen CPU & GoogLeNet & 96 & NHWC \\ \hline
\end{tabular}
\caption{Models used for evaluation}
\label{table:tfcnn_config}
\vspace{-0.1in}
\end{figure}

\begin{figure}
\centering
\begin{tabular}{||l|l|l|l|l||}
\hline
& \texttt{int} & \texttt{int} & \texttt{OMP\_} & \texttt{KMP\_} \\
& \texttt{er\_} & \texttt{ra\_} & \texttt{NUM\_} & \texttt{BLO} \\
\textbf{Model} & \texttt{op} & \texttt{op} & \texttt{THR} & \texttt{CK\_} \\
& & & \texttt{EADS} & \texttt{TIME} \\ \hline \hline
ResNet-50 & 2 & 56 & 56 & 1 \\ \hline
Inception3 & 2 & 56 & 56 & 1 \\ \hline
VGG16 & 1 & 56 & 56 & 1 \\ \hline
VGG11 & 1 & 56 & 56 & 1 \\ \hline
GoogLeNet & 2 & 56 & 56 & 1 \\ \hline
\end{tabular}
\caption{Environment variables used for MKL backend evaluation}
\label{table:mkl_tfcnn_config}
\vspace{-0.1in}
\end{figure}

\paragraph{\bfit{Configurations for parameter search space}} As mentioned in the
problem formulation section (section~\ref{sec:design}), we add strict upper and
lower bounds on the parameters to restrict the search space.
Figure~\ref{table:param_search} specifies the bounds used for our evaluation.
These values are derived from the number of cores available on the machine that
we used for evaluation (56 physical cores). But one can choose these ranges and
step sizes depending on their needs --- it is not necessary to stick to these
ranges.

\begin{figure}
\centering
\begin{tabular}{||l|l|l|l||}
\hline
& \texttt{inter\_} & \texttt{intra\_} & \texttt{OMP\_} \\
\textbf{Backend} & \texttt{op} & \texttt{op} & \texttt{NUM\_} \\
& & & \texttt{THREADS} \\ \hline \hline
MKL & [1, 4, 1] & [14, 56, 7] & [14, 56, 7] \\ \hline
Eigen & [1, 4, 1] & [14, 56, 7] & - \\ \hline
\end{tabular}
\caption{[Lower bound, upper bound, step size] for parameter search}
\label{table:param_search}
\vspace{-0.15in}
\end{figure}

\begin{figure*}[!ht]
\centering
\begin{subfigure}{0.5\textwidth}
  \includegraphics[width=\linewidth]{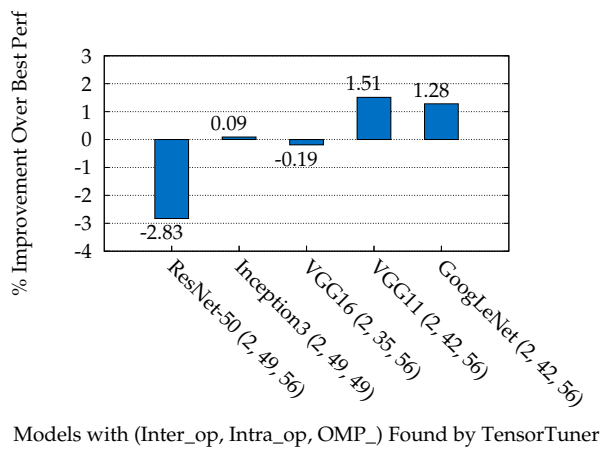}
	\vspace{-0.35in}
	\caption{Tuning Quality of MKL Backend for Training}
  \label{fig:mkl_train_quality}
\end{subfigure}%
\begin{subfigure}{0.5\textwidth}
  \includegraphics[width=\linewidth]{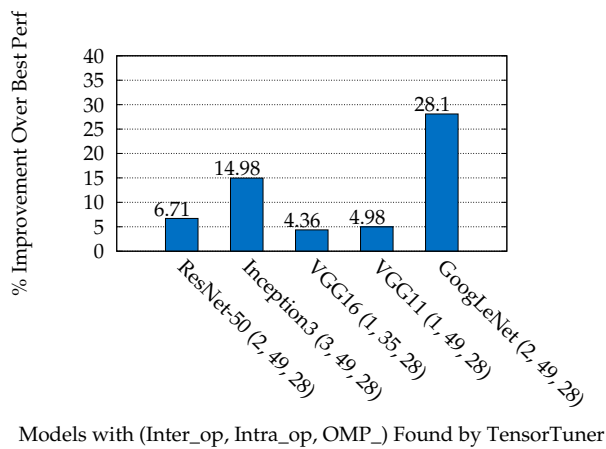}
	\vspace{-0.35in}
	\caption{Tuning Quality of MKL Backend for Inference}
  \label{fig:mkl_infer_quality}
\end{subfigure} %

\begin{subfigure}{0.5\textwidth}
  \includegraphics[width=\linewidth]{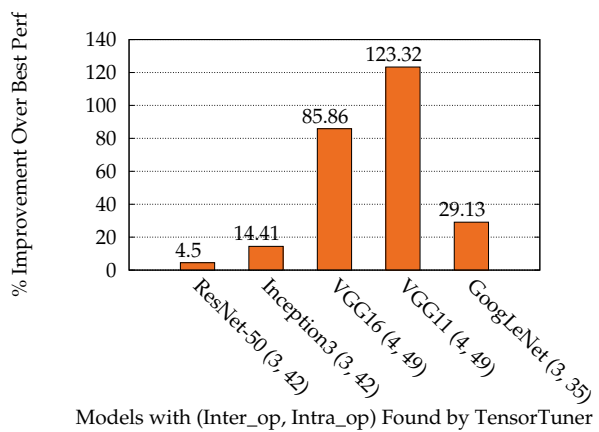}
	\vspace{-0.35in}
	\caption{Tuning Quality of Eigen Backend for Training}
  \label{fig:eigen_train_quality}
\end{subfigure}%
\begin{subfigure}{0.5\textwidth}
  \includegraphics[width=\linewidth]{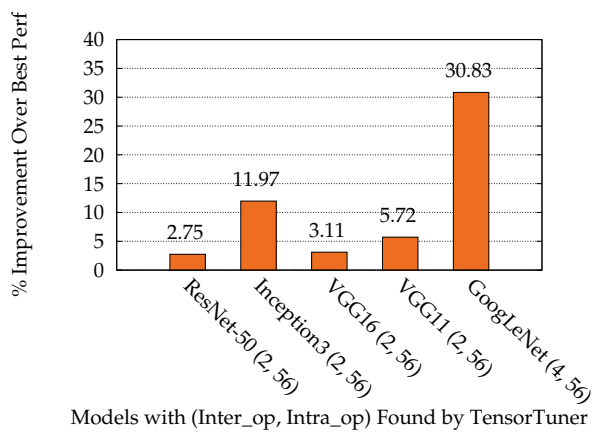}
	\vspace{-0.35in}
	\caption{Tuning Quality of Eigen Backend for Inference}
  \label{fig:eigen_infer_quality}
\end{subfigure}
\caption{{\ourapproach} Tuning Quality on MKL and Eigen CPU Backends}
\label{fig:tuning_quality}
\vspace{-0.15in}
\end{figure*}

\subsection{Tuning quality}

Tuning quality evaluates {\ourapproach}'s ability to find global optimum
parameter settings. Concretely, we evaluate tuning quality by comparing the
performance score that we get with the optimal parameter setting suggested by
{\ourapproach} with the score obtained using the best-known parameter setting. All
the best-known parameter settings for TensorFlow's MKL CPU backend are mentioned
in the Intel's recent blog \cite{inteltfv2}. For Eigen CPU backend, TensorFlow's
performance guide \cite{tfbenchmarkingguide} mentions that the default parameter
settings in TensorFlow's source code (for
\texttt{inter\_op\_parallelism\_threads} and
\texttt{intra\_op\_parallelism\_threads}) are efficient for the most systems.

\camerareadyonly{
\fancyhf{}
\thispagestyle{fancy}
\renewcommand{\headrulewidth}{0pt}
\lfoot{\vspace{-0.25in}\small{\textbf{DISCLAIMER}: Tests document performance of
components on a particular test, in specific systems. Differences in hardware,
software, or configuration will affect actual performance. Consult other sources
of information to evaluate performance as you consider your purchase.  For more
complete information about performance and benchmark results, visit
www.intel.com/benchmarks. Intel technologies` features and benefits depend on
system configuration and may require enabled hardware, software or service
activation. Performance varies depending on system configuration. No computer
system can be absolutely secure. Check with your system manufacturer or retailer
or learn more at \href{www.intel.com}{www.intel.com}.}}
}

\paragraph{\bfit{Evaluating tuning quality on TensorFlow's MKL CPU backend}}
Figure~\ref{fig:mkl_train_quality} compares the training performance of
TensorFlow's MKL CPU backend that is tuned using the best-known parameter
settings (baseline) with that obtained using the optimal parameter settings
found by {\ourapproach}.  Labels on the X-axis show the parameter settings
suggested by {\ourapproach} and the Y-axis shows the percentage improvement that
we got with {\ourapproach} suggested settings over baseline. For all the
topologies {\ourapproach} could find a setting that delivered the performance
close to the best-known performance from TensorFlow's MKL CPU backend. In fact,
for GoogleNet, the settings found by {\ourapproach} produced better performance
(by 1.28\%) than the best-known performance from Intel's blog!

Figure~\ref{fig:mkl_infer_quality} shows the inference performance of
TensorFlow's MKL CPU backend with the settings found by {\ourapproach} and
compares that with the performance obtained using the best-known settings
(baseline). Similar to the training plot, labels on the X-axis show the
parameter settings. Continuing the story from training results, inference
results also show that {\ourapproach} could find settings that produced better
performance than the performance produced by the best-known settings on all the
topologies! To be precise, settings found by {\ourapproach} produced 1.28\% to
28\% (28\% improvement on GoogLeNet) improved performance than the performance
produced by the best-known settings.  This points to the fact that manual
tuning, if it is not systematic, can lead to sub-optimal parameter settings.

In order to understand the effectiveness of {\ourapproach} in its ability to
converge to global optimum, we performed an exhaustive evaluation by scanning
the whole parameter search space for InceptionV3 training run and obtained the
performance score for every parameter setting. The exhaustive search found the
settings (2, 56, 49) that delivered 1.47\% better performance than the optimal
settings found by {\ourapproach} (2, 49, 49 as per
Fig~\ref{fig:mkl_train_quality}). This result highlights the ability of
{\ourapproach} and the underlying Nelder-Mead Simplex algorithm to get close to
the global optimum, if not converge to it. Nonetheless, it is interesting to
understand why {\ourapproach} could not find the best setting for InceptionV3,
and if we can tune the Nelder-Mead algorithm to find that setting.  Various
settings of Nelder-Mead such as radius used to construct initial simplex and
convergence criteria could make a difference. We plan to do this analysis as a
part of future work.

\paragraph{\bfit{Evaluating tuning quality on TensorFlow's Eigen backend}}

\camerareadyonly{
\fancyhf{}
\thispagestyle{fancy}
\renewcommand{\headrulewidth}{0pt}
\lfoot{\small{\textbf{DISCLAIMER}: The benchmark results may need to be
revised as additional testing is conducted.  The results
depend on the specific platform configurations and workloads utilized in the
testing, and may not be applicable to any particular user's components, computer
system or workloads. The results are not necessarily representative of other
benchmarks and other benchmark results may show greater or lesser impact from
mitigations.}}
}

After the evaluation of TensorFlow's MKL backend, we evaluated its Eigen CPU
backend for training and inference scenarios. Since TensorFlow sets the default
parameter values for Eigen backend statically, this is an evaluation of the
effectiveness of such a static approach and its generality.

Figure~\ref{fig:eigen_train_quality} compares the training performance obtained
using the default settings\footnote{To trigger the default settings for
\texttt{inter\_op\_parallelism\_threads} and
\texttt{intra\_op\_parallelism\_threads}, we set both of them to 0 as the
command line arguments to \texttt{tf\_cnn\_benchmarks}.} with that obtained using
the optimal settings found by {\ourapproach}. Similar to earlier figures, labels
on the X-axis indicate the parameter settings used to obtain the performance
improvements that are plotted on the Y-axis. Note that the settings found by
{\ourapproach} delivered 4\% to 123\% improvement in the training performance
over the performance obtained using the default settings.
Figure~\ref{fig:eigen_infer_quality} compares the inference performance for the
same set of models. Even for inference scenario, the optimal settings found by
{\ourapproach} delivered 2\% to 30\% improvement in the performance over the
performance obtained using the default settings.  This clearly highlights the
fact that the static approach for default parameter settings in TensorFlow's
Eigen CPU backend is not an ideal approach.

\paragraph{\bfit{Comparison of settings found by {\ourapproach} with the default
settings}}

\camerareadyonly{
\fancyhf{}
\thispagestyle{fancy}
\renewcommand{\headrulewidth}{0pt}
\lfoot{\small{\textbf{DISCLAIMER}: \emph{Configurations:} CentOS 7.3.1611 operating system
used in the evaluation was patched for GPZ variants 1 (CVE-2017-5753), 2
(CSE-2017-5715), and 3 (CVE-2017-5754) as of 1st of April, the date of
experimentation (Variants 1 and 2 are also known as Spectre, and variant 3 is
known as Meltdown). However, it was not patched for GPZ variants 3a and 4, which
were discovered after the experiments were performed.}}
}

\begin{figure}[!ht]
\centering

\begin{subfigure}{0.5\textwidth}
  \includegraphics[width=\linewidth]{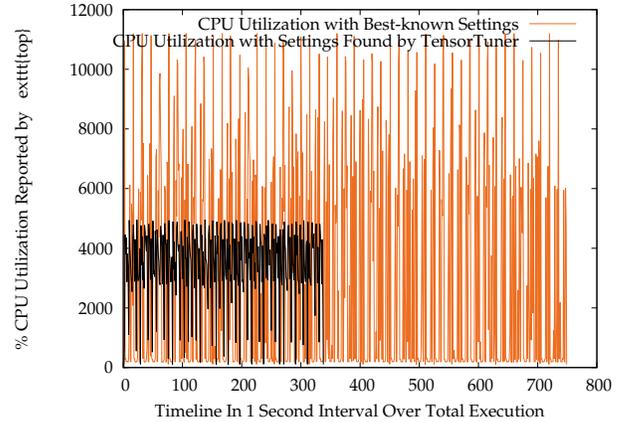}
	\vspace{-0.25in}
	\caption{CPU Utilization of VGG11 with Eigen CPU Backend}
  \label{fig:vgg11_cpu_usage}
\end{subfigure}%

\begin{subfigure}{0.5\textwidth}
  \includegraphics[width=\linewidth]{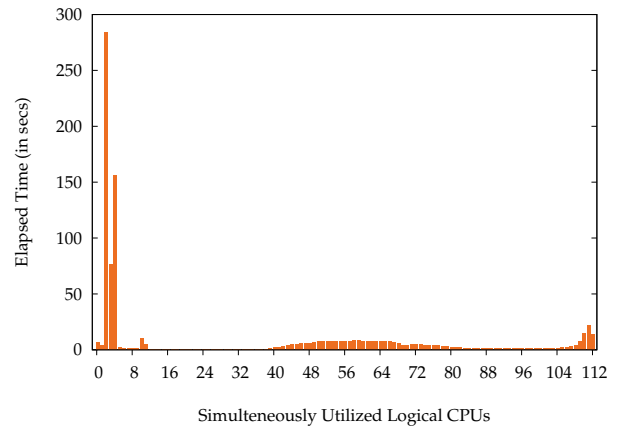}
	\vspace{-0.25in}
	\caption{CPU Utilization With Default Settings}
  \label{fig:cpu_usage_inter0_intra0}
\end{subfigure}%

\begin{subfigure}{0.5\textwidth}
  \includegraphics[width=\linewidth]{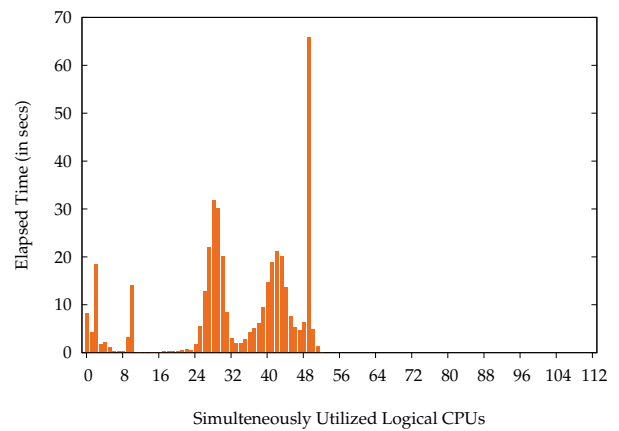}
	\vspace{-0.25in}
	\caption{CPU Utilization With {\ourapproach} Found Settings}
  \label{fig:cpu_usage_inter4_intra49}
\end{subfigure}

\caption{Analysis of VGG11 Performance with Eigen CPU Backend}
\label{fig:settings_comparison}
\vspace{-0.25in}
\end{figure}

The settings found by {\ourapproach} delivered 123\% improvement in VGG11
training performance with Eigen backend over the default settings. To get more
insights into this performance improvement, we collected CPU utilization over
the whole duration of VGG11's training run. The utilization was obtained for
every 1 second from Linux \texttt{top} utility. Figure~\ref{fig:vgg11_cpu_usage}
shows the CPU utilization for the training run with the settings found by
{\ourapproach} in black color, while the CPU utilization for the training run
with the default settings is shown in orange color. Notice that the complete
training run with the default settings took 751 seconds, while it took 340
seconds with the settings found by {\ourapproach}. Also notice that the peak CPU
utilization for the run with the default settings is almost 11200\%. This is
because {\intel} {\xeon} Platinum 8180 processor has 28 physical cores in 1
socket and 56 physical cores with 2 sockets. Additionally, with hyper-threading
enabled (and 2 threads per core), the total number of cores in a 2-socket 8180
processor is 112. The most important observation though is that the average CPU
utilization during the run with the settings found by {\ourapproach} was much
less than during the run with the default settings. This points to thread
over-subscription issue.  The difference between average CPU utilizations was
also recorded by the report obtained from Intel's VTune Amplifier
(Figure~\ref{fig:cpu_usage_inter0_intra0} and
Figure~\ref{fig:cpu_usage_inter4_intra49}).

To confirm the thread over-subscription issue, we repeated the training
experiment with the same settings but by exposing only 56 CPU cores to the
TensorFlow framework\footnote{The number of CPUs were restricted using
\texttt{numactl} utility.}. We saw that the VGG11 training performance with the
default settings improved by almost 26\%. In other words, the performance gap
was reduced from 123\% to 65\%. The report obtained using Intel's VTune
Amplifier also confirmed the reduction in the thread spin time, leading to
the performance improvement.

\subsection{Efficiency of {\ourapproach}}

\begin{figure*}[!ht]
\begin{center}
\begin{subfigure}{0.5\textwidth}
  \includegraphics[width=\linewidth]{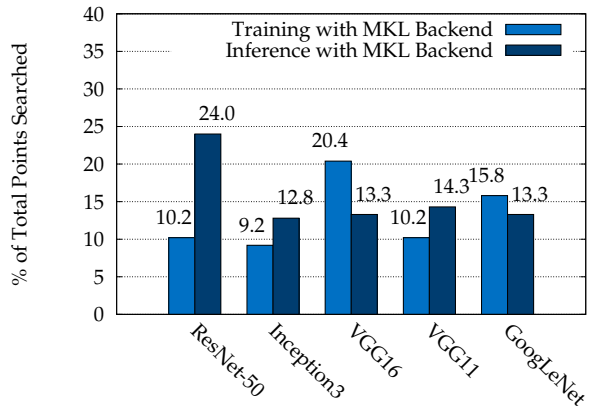}
	\vspace{-0.35in}
	\caption{Tuning Efficiency of MKL Backend}
  \label{fig:mkl_tuning_efficiency}
\end{subfigure}%
\begin{subfigure}{0.5\textwidth}
  \includegraphics[width=\linewidth]{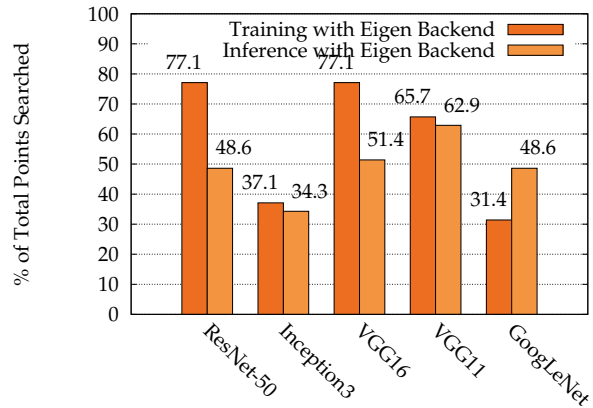}
	\vspace{-0.35in}
	\caption{Tuning Efficiency of Eigen Backend}
  \label{fig:eigen_tuning_efficiency}
\end{subfigure}%
\caption{{\ourapproach} Tuning Efficiency on MKL and Eigen CPU Backends}
\label{fig:tuning_efficiency}
\end{center}
\vspace{-0.25in}
\end{figure*}

We evaluate efficiency of our approach by comparing the number of points in the
parameter space searched by {\ourapproach} with the total number of points that
would be explored by an exhaustive search.
Figure~\ref{fig:mkl_tuning_efficiency} compares the efficiency
in terms of tuning the MKL CPU backend for training and inference modes.  In
comparison to 196 points that would be explored by an exhaustive search,
{\ourapproach} searches from 9\% to 24\% of the points to find the optimal
settings. It could be possible to further reduce the number of points searched
by tuning the expansion or reduction settings of Nelder-Mead algorithm, but that
could also affect the tuning quality.

Figure~\ref{fig:eigen_tuning_efficiency} compares the efficiency of
{\ourapproach} for tuning Eigen CPU backend for training and inference modes.
In comparison to the 35 points that would be explored by an exhaustive search,
{\ourapproach} searches from 31\% to 77\% of the points to find the optimal
settings. More than 50\% of the points searched by {\ourapproach} could
potentially be because of the smaller search space that reduces the
effectiveness of Nelder-Mead algorithm in quickly pruning the search space.

\section{Related work}
\label{sec:relwork}

\paragraph{\bfit{Optimization algorithms}}
The problem of auto-tuning TensorFlow's CPU backend comes under a general
class of \emph{mathematical optimization problems}. Considerable research over multiple
centuries has been devoted to the problem of optimizing linear functions,
non-linear functions, single-variable and multi-variables functions, functions
with constraint and without constraints, convex and non-convex functions. Of
course, the discussion here cannot cover all of the past research, so an
interested reader can refer to various books \cite{book1} on mathematical
optimizations.

The problem of auto-tuning TensorFlow's CPU backend can be formulated as a
\emph{black-box optimization problem} or a \emph{white-box optimization
problem}.  Black-box optimization problems are those in which the objective
function $f : X \rightarrow \mathbb{R}$ can only be evaluated ($f(x)$) for any $x \in
X$ but we have no access to any other information about $f$ such as gradients or
the Hessian.  White-box optimization problems, on the other hand, are those in
which the additional information about $f$ is available.

White-box optimization problems can be efficiently solved using
\emph{gradient-based function optimization algorithms} as the gradients or
Hessian are available. These algorithms (such as gradient descent, Newton's
method, Quasi-Newton methods \cite{quasinewton} such as DFP method and BFGS
method) use additional information available about the function to decide the
promising directions along which to search for an optimum solution. For a large
number of variables, gradient-based optimization algorithms are typically the
most efficient algorithms.  Unfortunately, gradient-based optimization
algorithms are inefficient at optimizing noisy, discontinuous, and non-convex
functions. Additionally, they do not support function parameters that are
discrete or mixed discrete-continuous.

Black-box optimization problems, on the other hand, can only be solved using
\emph{gradient-free function optimization algorithms}. A number of gradient-free
optimization algorithms such as Nelder-Mead Simplex \cite{NM}, Simulated
Annealing, Divided Rectangles method, Genetic Algorithms and Particle Swarm
Optimization \cite{optimization_book} exist but have their own weaknesses.
Gradient-free optimization algorithms are typically expensive when the objective
function has a large number of parameters. More importantly, gradient-free
optimization algorithms are not guaranteed for find global optimum as the
problem of finding global optimum is NP-complete (since they need to evaluate
$f(x)$ $\forall$ $x$ $\in$ $X$ in order to find global optimum.)

\paragraph{\bfit{Auto-tuning in high-performance computing systems}} Automatic
tuning of parameters for best performance is a well-researched area in high
performance computing (HPC) domain \cite{bailey2010performance}.  Performance is
typically the most important objective function in many scientific and HPC
applications.  Unsurprisingly, considerable efforts has been devoted to solving
this problem. Since matrix multiplication is the one of the fundamental
computations in many HPC applications, considerable research efforts were
devoted to auto-tuning matrix multiplication kernels \cite{whaley,
vuduc2005oski}.  Given the raw compute power of GPUs, considerable research
efforts have focused on performance optimization for GPUs as well
\cite{li2009note, grauer2012auto}.

In machine learning domain, auto-tuning is routinely applied to the problem of
\emph{hyper-parameter tuning} (e.g., HyperOpt \cite{hyperopt},
MOE\footnote{\href{https://github.com/Yelp/MOE}{https://github.com/Yelp/MOE}},
Spearmint \cite{spearmint}, AutoWeka \cite{autoweka, autowekav2}, and
Hypertune\footnote{
\href{https://cloud.google.com/ml/}{https://cloud.google.com/ml/}} subsystem in
Google Cloud Machine Learning Engine that uses Google's Vizier \cite{vizier})
and \emph{automatic generation of efficient kernels of neural network
operations} (e.g., Tensor Comprehension \cite{tensor_comprehension}).

Hyper-parameters are tunable parameters in neural networks. Typical
hyper-parameters such as batch size, learning rate, etc, are critical to
convergence as well as good training or inference accuracy. Hyper-parameter
tuning also takes considerable time, since neural networks typically take long
time to converge. It is unsurprising then to see research effort dedicated to
solving this problem. All of the existing hyper-parameter tuning techniques apply
to the problem of tuning TensorFlow's CPU backend, since the threading model can
be considered as a set of hyper-parameters. {\ourapproach}, on the other hand,
can also be applied to the tuning problems solved using existing hyper-parameter
tuning tools. All of these tools have fairly similar strengths --- ability to
define configurable search spaces, ability to plug-in new search algorithms
(such as Batched Gaussian Process Bandits, Tree-of-Parzen-Estimators (TPE),
random search, grid search, etc), ability to support real-valued, integer-valued
or discrete-valued parameters, etc.

\blockcomment{
Google's Cloud Machine Learning Engine service used by developers and data
scientists to build machine learning models has \emph{Hypertune} subsystem for
auto-tuning model hyper-params for performance. Hypertune
subsystem uses Google's Vizier \cite{vizier} which is a Google-internal service
for performing black-box optimizations. At its core, Vizier supports closed
parameter search spaces over real values and integers.  Additionally, the
search space can also be explicitly specified as an ordered set of real numbers
or unordered set of strings. Vizier also supports a number of optimization
algorithm such as Batched Gaussian Process Bandits (default), random search,
grid search, etc.
}

We are not aware of any existing research work that applies auto-tuning
techniques to improve performance of TensorFlow's CPU backend on a neural
network model. TensorFlow's GPU backend, however, uses an auto-tuning technique
to choose the best convolution algorithm at the beginning of training or
inference of a neural network. But unlike {\ourapproach} that optimizes the
threading model parameters, the auto-tuning technique used by TensorFlow's GPU
backend seems to\footnote{No standard reference, other than TensorFlow's source
code, exists for this work} just scan the list of available convolution
algorithms at the start and pick up the best-performing one after trial
executions.  {\ourapproach}, in that sense, is a black-box optimization
approach.

\section{Future work}
\label{sec:future}

Although {\ourapproach} is able to find many parameter settings that deliver
better performance for TensorFlow's MKL and Eigen CPU backends over the
best-known performance numbers, there are a number of avenues for improvement.
First, although Nelder-Mead Simplex algorithm was able to find optimal parameter
settings efficiently, the question about the effectiveness of other
gradient-free search strategies is not answered. Additionally, Nelder-Mead
algorithm is known to have convergence issues with a large number of design
variables. So far we explored the algorithm with a maximum of 3 variables; it is
interesting to explore applicability of Nelder-Mead simplex algorithm to more
number of TensorFlow design variables. We evaluated {\ourapproach} on
TensorFlow's benchmark suite, and the applicability of the technique to general
models with large datasets, such as ImageNet, is unknown. Although the model
along with its dataset is a black-box for {\ourapproach}, the number of threads
used for pre-processing a dataset could affect the optimal settings for
TensorFlow's threading model.

\blockcomment{
We have upstreamed all of our optimizations; however, they do not represent the
end of our optimization efforts by any means. Newer deep learning models may
demand further optimizations. Moreover, {\intel} {\xeon} processor E5 v4
(codename Broadwell) and {\intel} {\xeonphi} processor 7250 (codename Knights
Landing) based platforms lay the foundation for next generation Intel products,
such as {\intel} {\xeon} (codename Skylake) and {\intel} {\xeonphi} (codename
Knights Mill). Although, MKL-DNN library already supports Skylake processors, we
may need some additional work to support these platforms in TensorFlow CPU backend.
In addition, multinode setup of TensorFlow models on Intel architecture will
demand further optimizations and tunings.
}

\section{Conclusion}
\label{sec:conclusion}
\thispagestyle{plain}

In this paper, we presented our approach, called {\ourapproach}, to the problem
of auto-tuning TensorFlow's threading model for MKL and Eigen CPU backends. Our
experimental evaluation on a set of popular convolution neural network models
from TensorFlow's benchmark suite revealed that the default parameter settings
used by Eigen CPU backend for both training and inference use case deliver
sub-optimal TensorFlow performance, and with the settings found by
{\ourapproach}, we get almost 2X improvement in performance. Even for the
well-tuned MKL CPU backend, we found that the publicly-available settings
delivered sub-optimal TensorFlow performance for inference use case (the
settings delivered optimal TensorFlow performance for training use case though,)
and that the settings found by {\ourapproach} delivered almost 30\% improvement
in performance for inference use case. Our experimental results underscore the
fact that the manual tuning of TensorFlow's CPU backends may not yield the best
TensorFlow performance, and an automated approach can tune the CPU backends much
better than the manual tuning.

\blockcomment{
In this paper, we presented our preliminary experimental results obtained while
optimizing TensorFlow deep learning framework on {\intel} {\xeon} and {\intel}
{\xeonphi} processors. Our evaluation results demonstrate effect of our
optimizations on default CPU backend in TensorFlow and deliver between 6X and
86X improvement. All of our optimizations are part of TensorFlow open-source
code now. This report does not mark the end of our optimization exercise; we
will be working on optimizing TensorFlow (including for future Intel
architectures) and will be upstreaming our optimizations along the way.
}

\section*{Acknowledgements}

We would like to thank anonymous reviewers for their comments and suggestions.
We would specifically like to thank Nagib Hakim for his insightful comments and
discussions on the preliminary version of the paper. Finally, we would also like
to thank all the members of Intel's TensorFlow optimization team for their
inputs for this work and for experimenting with {\ourapproach}.

\bibliographystyle{IEEEtran}
\bibliography{ms}

\appendix
\section{Appendix}
\label{sec:appendix}

In this section, we provide commands that we used for experiments. All the
configuration details used for experiments are provided in the Evaluation
section.

\vspace{0.1in}
\paragraph{\bfit{Command to run TensorTuner to tune MKL CPU backend for training}}

\begin{verbatim}
harmony/bin/tuna STRATEGY=nm.so -q -v
-i=interop,$interop_min,$interop_max,$interop_step
-i=intraop,$intraop_min,$intraop_max,$intraop_step
-i=omp,$omp_min,$omp_max,$omp_step
numactl -l python tf_cnn_benchmarks.py
--forward_only=False --num_warmup_batches=0
--batch_size=$batch_size --data_format=NCHW
--num_batches=100 --model=$model
--num_inter_threads=%interop
--num_intra_threads=%intraop
--num_omp_threads=%omp
\end{verbatim}

\vspace{0.1in}
\paragraph{\bfit{Command to run TensorTuner to tune Eigen CPU backend for training}}

\begin{verbatim}
harmony/bin/tuna STRATEGY=nm.so -q -v
-i=interop,$interop_min,$interop_max,$interop_step
-i=intraop,$intraop_min,$intraop_max,$intraop_step
python tf_cnn_benchmarks.py
--forward_only=False --num_warmup_batches=0
--batch_size=$batch_size --data_format=NHWC
--num_batches=100 --model=$model
--num_inter_threads=%interop
--num_intra_threads=%intraop
\end{verbatim}

\vspace{0.1in}
\paragraph{\bfit{Command to run TensorTuner to tune MKL and Eigen CPU backends for
inference}}

Running their respective commands with \texttt{--forward\_only=True} enables inference mode
of \texttt{tf\_cnn\_benchmarks.py}.

\end{document}